 \documentclass{emulateapj}
\usepackage{natbib}

  \shortauthors{Roques et al.}
  \shorttitle{High Energy Emission of V404 Cygni during 2015 outburst with \textit{INTEGRAL}/SPI}

\begin{document}

\title{High Energy Emission of V404 Cygni during 2015 outburst with \textit{INTEGRAL}\footnotemark[1]/SPI: Spectral analysis issues and solutions
}
\footnotetext[1]{Based on observations with \textit{INTEGRAL}, an ESA project with instruments and science data centre funded by ESA member states
 (especially the PI countries: Denmark, France, Germany, Italy, Spain, and Switzerland), Czech Republic and Poland with participation of
 Russia and USA.}
 
\author{Jean-Pierre~Roques\altaffilmark{1},Elisabeth~Jourdain\altaffilmark{1}\\
}

%\altaffiltext{1}{Universit\'e de Toulouse; UPS-OMP; IRAP;  Toulouse, France\\
% CNRS; IRAP; 9 Av. colonel Roche, BP 44346, F-31028 Toulouse cedex 4, France}
%\altaffiltext{2}{Istituto di Astrofisica e Planetologia Spaziali, INAF, Via Fosso del Cavaliere 100, Roma, I-00133, Italy}

\affil{\textsuperscript{1}Universit\'e Toulouse; UPS-OMP; CNRS; IRAP; 9 Av. Roche, BP 44346, F-31028 Toulouse, France\\
}
%\author{\it Received ; accepted  ...  }
%\altaffiltext{2}{{\it To whom proofs and offprint requests should be sent}
%\email{jourdain@cesr.fr}}

\begin{abstract}
A strong outburst of the X-ray transient V404 Cygni (= GS2023-338) was observed in 2015 June/July  up to a level of 50 Crab in the hard X-ray domain. At this level of photon flux, an instrument's behavior may be severely tested and some instrumental artifacts could affect the data analysis.
We are interested in the SPI instrument aboard the INTEGRAL mission and have performed thorough checks to ensure a correct handling of the data. By analyzing the observations throughout the outburst, we have observed that the high energy domain (above 500 keV) sometimes exhibits  unexpected features which are worth careful examination. Spurious triggers are known to affect the MeV region and we suspected that this phenomenon could be accentuated by the huge photon flux. We have investigated this issue, specifically during high flux periods and actually found that artificial high energy bumps may appear with the current standard analysis procedure.  However, 
if the specific selection events usually used in the 650-2200 keV energy  is applied down to 450 keV,   
  the spurious noise and the associated spectral features are removed. We present how to obtain reliable spectral results on the high energy emission of V404 Cyg at extreme flux levels and demonstrate that with the correct configuration, the hard X-ray emission, up to a few MeV, is  modeled by a two component model as observed in Cyg X-1 and for V404 Cygni itself at lower flux levels. 

\end{abstract}      


\maketitle

%%%%%%%%%%%%%%%%%%%%%%%%%%%%%%%%%%%%%%%%%%%%%%%%%%%%%%%%%%%%%%%%%%%%%%%%%
\section{Introduction}
%%%%%%%%%%%%%%%%%%%%%%%%%%%%%%%%%%%%%%%%%%%%%%%%%%%%%%%%%%%%%%%%%%%%%%%%%

In June 2015, the source V404 Cygni (also called GS2023-338) underwent an extraordinary outburst (Barthelmy et al. 2015, Younes, 2015) reaching never observed flux levels in hard X-rays. The INTEGRAL observatory was pointed on the source two days after the alert for a campaign of almost four weeks of continuous monitoring. The source exhibited particularly intense flares in the first twelve days, corresponding to INTEGRAL revolution numbers 1554 to 1557. High energy emission during the first three days (revolution 1554, immediately public data) has been investigated with both SPI and IBIS instruments in two papers (Roques et al., 2015 and Natalucci et al. 2015). During this first revolution, the (highly variable) source intensity varied mainly between half and ten or so Crab, and the spectral shape has been described with two components corresponding to a Comptonisation  emission plus an additional tail extending up to $\sim$ 750 keV as a cutoff powerlaw.  Later, the source activity still increased with extreme amplitude variability on all timescales and several intense flares that reached  brightness levels up to 50 Crab. Diehl et al. (2015) suggest that spectral features, possibly related to positron annihilation, are intermittently present.
 However, such intense fluxes may be a problem for instruments. Significant losses of telemetry or an increase of the dead time are the most obvious but electronic system limits may also be revealed. 
 It is thus important to understand the instrument behavior to ensure a reliable procedure to analyze the data, especially during such exceptional conditions. 

 In this paper, we first give a short description of the SPI instrument, focusing on the electronics since it is crucial to understanding the behavior of the recorded signal.  We then present the data analysis of a sample of observations, at different source flux levels to illustrate potential instrumental effects. Finally, we describe the simple, tested procedure to be implemented  to assess the reliability of the final scientific results in such cases of exceptionally high photon fluxes.

%%%%%%%%%%%%%%%%%%%%%%%%%%%%%%%%%%%%%%%%%%%%%%%%%%%%%%%%%%%%%%%%%%%%%%%%%
\section{Instrument, observations and data analysis}\label{instru}
%%%%%%%%%%%%%%%%%%%%%%%%%%%%%%%%%%%%%%%%%%%%%%%%%%%%%%%%%%%%%%%%%%%%%%%%
%%%%%%%%%%%%%%%%%%%%%%%%%%%%%%%%%%%%%%%%%%%%%%%%%%%%%%%%%%%%%%%%%%%%%%%%%
\subsection{Brief description of the SPI spectrometer}
%%%%%%%%%%%%%%%%%%%%%%%%%%%%%%%%%%%%%%%%%%%%%%%%%%%%%%%%%%%%%%%%%%%%%%%% 
 The SPI spectrometer is one of the two main instruments aboard the INTEGRAL mission launched in October 2002. The detection plane is made of 19 crystals of high purity Germanium maintained at 80 K to ensure an energy resolution between 2 and 8 keV in the energy domain from 20 to 8000 keV. Due to a coded mask located 1.8 m above the detection plan, SPI provides images of the sky with a spatial resolution of 2.2$^\circ$ over a 30$^\circ$ field of view. An anticoincidence system surrounding the camera detects the particles arriving from outside the field of view, to measure the ambient background and remove it from the signal. More details can be found in Winkler et al. (2003) for the INTEGRAL mission and in Vedrenne et al. (2003) and Roques et al. (2003) for the SPI instrument.
 %%%%%%%%%%%%%%%%%%%%%%%%%%%%%%%%%%%%%%%%%%%%%%%%%%%%%%%%%%%%%%%%%%%%%%%%%
\subsection{The electronics and spurious triggers}
%%%%%%%%%%%%%%%%%%%%%%%%%%%%%%%%%%%%%%%%%%%%%%%%%%%%%%%%%%%%%%%%%%%%%%%%
 Initial checks performed at the very beginning of the mission have shown the presence of a variable, unexpected signal in the MeV region, which has been understood in terms of spurious triggers, probably linked to saturating particles, and generating "false events" in the main electronic chain (see for instance http://sigma-2.cesr.fr/integral/meetings/4303-CR\_SPI-CoIs\_March17.pdf for a first report on that phenomenon). However, the signal produced by the photons interacting with the detector plane are analyzed by two independent electronic chains. The first one provides energy and time information for all the events recorded in the detector plane. A second electronic analyzer was dedicated to the Pulse Shape Discrimination (PSD). These electronics analyze  the fast output  of the preamplifier (ie the image of the current pulse produced by the interaction in the Germanium detectors). They are much faster and  so recover much faster after a saturating event. Due to its specificity, the PSD electronic chain  operates in an energy domain restricted to 450 keV- 2.2 MeV (note that the lower threshold has been changed during the mission, the value given here corresponds to the period of interest). The events analyzed by the PSD channel are identified by a PSD flag, and we have to take into account that its efficiency is $\sim$ 15\% less than the main electronic chain.

During the investigations around the spurious noise observed around 1 MeV, various tests have been conducted. In particular, the analysis has been restricted to the events seen by both main and PSD electronic chains (PSD flagged). The main conclusion is that this subset of events is not polluted by the spurious emission and provides a clean signal. Thus, a standard analysis procedure has been proposed (Jourdain and Roques, 2009): In the energy band 650-2200 keV, only the events idenitified with the PSD flag are extracted  from  the data (PSD events). A correction factor of 1/0.85 is applied to compensate the efficiency loss. Below 650 keV (and above 2.2 MeV), all the events are considered. This procedure has been successfully tested during observations of bright hard sources like Crab Nebula and Cyg X-1. 
Note that photons which interact in one detector and are absorbed in another one, (called 'multiple events' (ME)) 
are not affected by this issue since their detection requires at least two independent simultaneous triggers in differents electronic chains. However, we will not consider them hereafter. 
 %%%%%%%%%%%%%%%%%%%%%%%%%%%%%%%%%%%%%%%%%%%%%%%%%%%%%%%%%%%%%%%%%%%%%%%%%
\subsection{Observations and Data analysis}
%%%%%%%%%%%%%%%%%%%%%%%%%%%%%%%%%%%%%%%%%%%%%%%%%%%%%%%%%%%%%%%%%%%%%%%% 
INTEGRAL observations of V404 Cyg consists of ~3 day revolutions, split into 50 exposures or science window(scw) of 3400 s, with pointing direction shifted by 2.2 $^{\circ}$ from one scw to the next one, to follow a 5X5 rectangular or a 7 hexagonal dithering pattern. 
Due to the huge source flux, telemetry limits can be reached. For the SPI instrument, the saturation of the telemetry is limited to the source maximum flux period (peak around 2015 June 26, 16:30 or MJD 57199.19)  where loss of telemetry packets becomes significant. It has to be taken into account properly for any timing study, but concerning spectral analyses, it only causes an increase in the deadtime from  less than 15\% (usual value due to electronic deadtime) to less than  25 \% ($\sim$ 10\% added due to telemetry gaps).
We have produced the count spectra of V404 Cyg for each scw by the standard sky fitting procedure. The sky model included Cyg X-1 and Cyg X-3 in addition to V404 Cyg. The background estimation is based on 'empty field' observations  during revolutions 1546 to 1549, with the same PSD selection criterion. The background amplitude has been allowed to vary on the scw timescale due to some significant solar activity. For each spectrum, fluxes are extracted in 43 logarithmically spaced energy bins from 20 keV to 1 MeV,
and deconvolved with the associated response through the {\sc xspec} (v12.8.2) tools.
Below 25 keV, uncertainties on the instrument responses may lead to significant deviations in the spectral residuals for any of the tested standard spectral models, and we ignored them in the presented results.

As mentioned above, the current standard analysis is based on the selection of PSD flagged events in the 650-2200 keV domain to avoid potential artefacts known to be present in the sub-MeV region. However, for test purposes, the selection criterion has been extended to photons between 450-650 keV. The efficiency of the PSD electronic chain has been measured by means of background lines with energy falling in the appropriate domain and appears to be very stable with time and with energy. A correction factor of 0.85 is thus applied during the data treatment in the appropriate energy band.

%%%%%%%%%%%%%%%%%%%%%%%%%%%%%%%%%%%%%%%%%%%%%%%%%%%%%%%%%%%%%%%%%%%%%%%%%
\section{Observations and Results}\label{results}
%%%%%%%%%%%%%%%%%%%%%%%%%%%%%%%%%%%%%%%%%%%%%%%%%%%%%%%%%%%%%%%%%%%%%%%%% 
In this section, we present results obtained for a sample of representative spectra, focusing more particularly on the periods where the source flux is highly variable and / or exhibits bright flares. We have selected three scws and the mean spectrum averaged over the five revolutions. Table 1 below gives the corresponding time intervals and durations.  We have performed two analyses, with and without selecting the PSD flagged events.

The results of these contradictory analyses are displayed in Fig. 1-4 for four periods presenting differents flux levels (3 scw intervals and the total averaged spectrum). While the results should be compatible if all the events detected between 450 keV and 2.2 MeV were real, it appears clearly that it is not the case. When the source flux is particularly high, a feature sometimes appears in the first analysis (blue points in the figures) while the selection of PSD flagged events (red points) drastically reduces the emission in this energy band.  In other words, the 500-1000 keV  energy range may be strongly affected by an instrumental artifact, probably due to the electronic spurious trigger effect described above and enhanced by the huge intensity of the source, producing a broad feature which vanishes when a  more cautious analysis is performed. 
 
As a scientific result, we point out that all the analyzed spectra can be described with only two spectral components, one
identified with a thermal Comptonisation (kT $\sim$ 20 -30 keV, $\tau \sim$ 0.5 -1.5) and the second one modeled with a cutoff power-law to account for the emission at high energy. This scenario is remarkably similar to that obtained in the first analyses dedicated to the first 3 days of the INTEGRAL campaign (2-5 days after the Swift alert; Roques et al. 2015) when the source was less active. It suggests that all along this outburst, the hard X-ray emission of V404 Cyg can be understood in terms of two spectral components, varying independently in shape (implying, for example, a change in the electron population temperature and optical depth) and in amplitude. No reliable spectral feature has been found in excess of these two components.

%%%%%%%%%%%%%%%%%%%%%%%%%%%%%%%%%%%%%%%%%%%%%%%%%%%%%%%%%%%%%%%%%%%%%%%%%
\section{summary and conclusion}
%%%%%%%%%%%%%%%%%%%%%%%%%%%%%%%%%%%%%%%%%%%%%%%%%%%%%%%%%%%%%%%%%%%%%%%%%
We have  analyzed the INTEGRAL/SPI  observations dedicated to the very bright outburst of the X-ray transient V404 Cygni.
The huge photon fluxes emitted by this source pushed us to further investigate the instrument behavior.  Considering a known instrumental effect as possibly responsible of spurious spectral features, we have tested a simple procedure consisting of analysis only PSD flagged events (so 85$\%$ of the total events recorded on the detector plan). While removing 15 $\%$ of the events should only slightly affect the resulting  spectra (in counts or in photons), we notice that some features present in the first analysis almost disappear when the PSD flag criterion is used down to 450 keV. By comparing both results, it is clear that features sometimes present  at high energy, particularly during flares, are essentially due to an enhanced electronic noise. The appropriate method to clean them has been detailed: it consists in using the PSD flagged events alone for E between 450 keV and 2.2 MeV. 

All the studied spectra provide statistically acceptable $\chi2$ values when fitted with a two component model similar to the Cyg X-1 emission. This corresponds to a scenario involving a Comptonizing electron population with a temperature of 20-30 keV and an optical depth around 1 plus an additional cutoff powerlaw, presenting a cutoff around 200-300 keV for an index power law fixed to 1.6.

In conclusion, we want to point out that the SPI data analysis can be safely performed with the standard procedure presented above.  
A tool dedicated to the SPI data spectral analysis with the proper configuration is available on http://sigma-2.cesr.fr/integral/spidai.
 
\section*{Acknowledgments}  The \textit{INTEGRAL} SPI project has been completed 
under the responsibility and leadership of CNES.

%%%%%%%%%%%%%%%%%%%%%%%%%%%%%%%%%%%%%%%%%%%%%%%%%%%%%%%%%%%%%%%%%%%%%%%%%%%%%
%%%% Table 1 %%%%%%%%%%%%%%%%%%%%%%%%%%%%%%%%%%%%%%%%%%%%%%%%%%%%%%%%%%%%%%%%
%%%%%%%%%%%%%%%%%%%%%%%%%%%%%%%%%%%%%%%%%%%%%%%%%%%%%%%%%%%%%%%%%%%%%%%%%%%%%
\begin{deluxetable}{lcccc}%1
\tablewidth{0pt}
\tablecaption{ \textit{INTEGRAL} SPI observations of V404 Cygni (=GS2023 -338) displayed in Figures 1 to 4.}
\label{tab:revol}
\tabletypesize{\scriptsize}
\tablehead{
 \colhead{}\\
\colhead{revol and scw}
&\colhead{Start} 
&\colhead{End} 
&\colhead{useful}\\
\colhead{number}
&\colhead{UT}
&\colhead{UT}
&\colhead{duration (ks)}\\
%&\colhead{} \\
 }
\startdata 
1555 -scw 19 & 2015-06-21 06:55:56 & 	2015-06-21 07:54:09  & 2.77        \\
1557 -scw 14 & 2015-06-26 12:09:33 &	2015-06-26 12:59:33    &  2.37   \\
1557 -scw 19 & 2015-06-26 16:29:47  &  2015-06-26 17:19:47    &  2.29  \\
1554 to 1559 & 2015-06-17 21:29:49  &2015-07-03 06:15:18 &  783.4       
\enddata
%\tablecomments{ }
\end{deluxetable}

%%%%%%%%%%%%%%%%%%%%%%%%%%%%%%%%%%%%%%%%%%%%%%%%%%%%
%%%%%%%%%%%  FIGURES %%%%%%%%%%%%%%%%%%%%%%

 \begin{figure}%1
%\plotone{8spectresV404.SPIBIS.ps} 
%\begin{figure}[t]
% \includegraphics[width=80mm, height= 80mm, angle=90,trim = 10mm  0mm 0mm 5mm, clip]{f2.ps}
%\end{figure}
 %\includegraphics[scale=0.4, angle=90,trim = 10mm  20mm 0mm 10mm, clip]{fig1.pdf}
%\plotone{V404withNoPSD.ps}
 \plotone{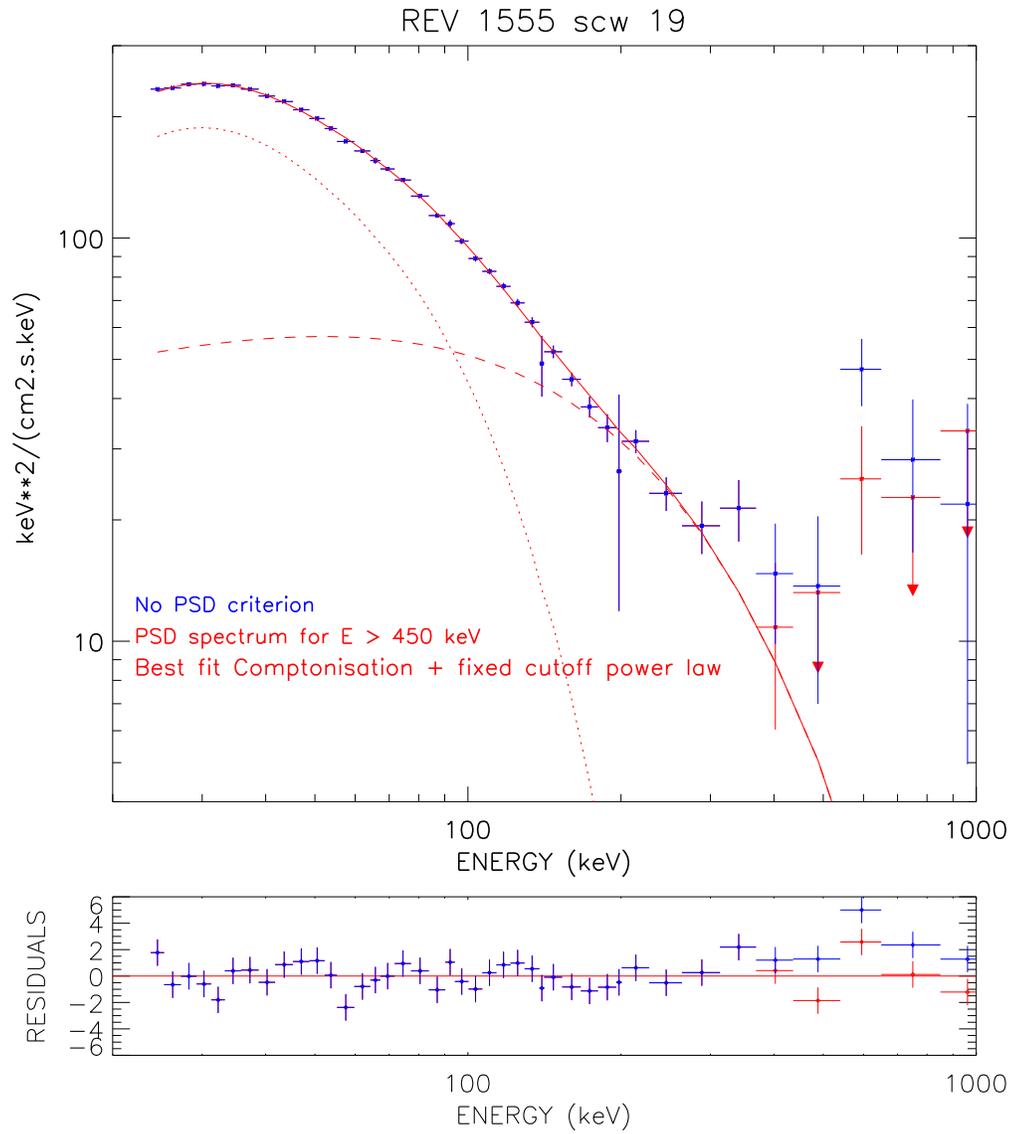}
 \caption{Comparison of the two data analysis procedures for the spectrum corresponding to the maximum of the revolution 1555; 'PSD spectrum' (red points) means that only PSD flagged events are taken into account in the analysis, i.e. events seen by two independent electronic chains.}
\label{fig:compascw1550-66}
\end{figure}

  \begin{figure}%1
 \plotone{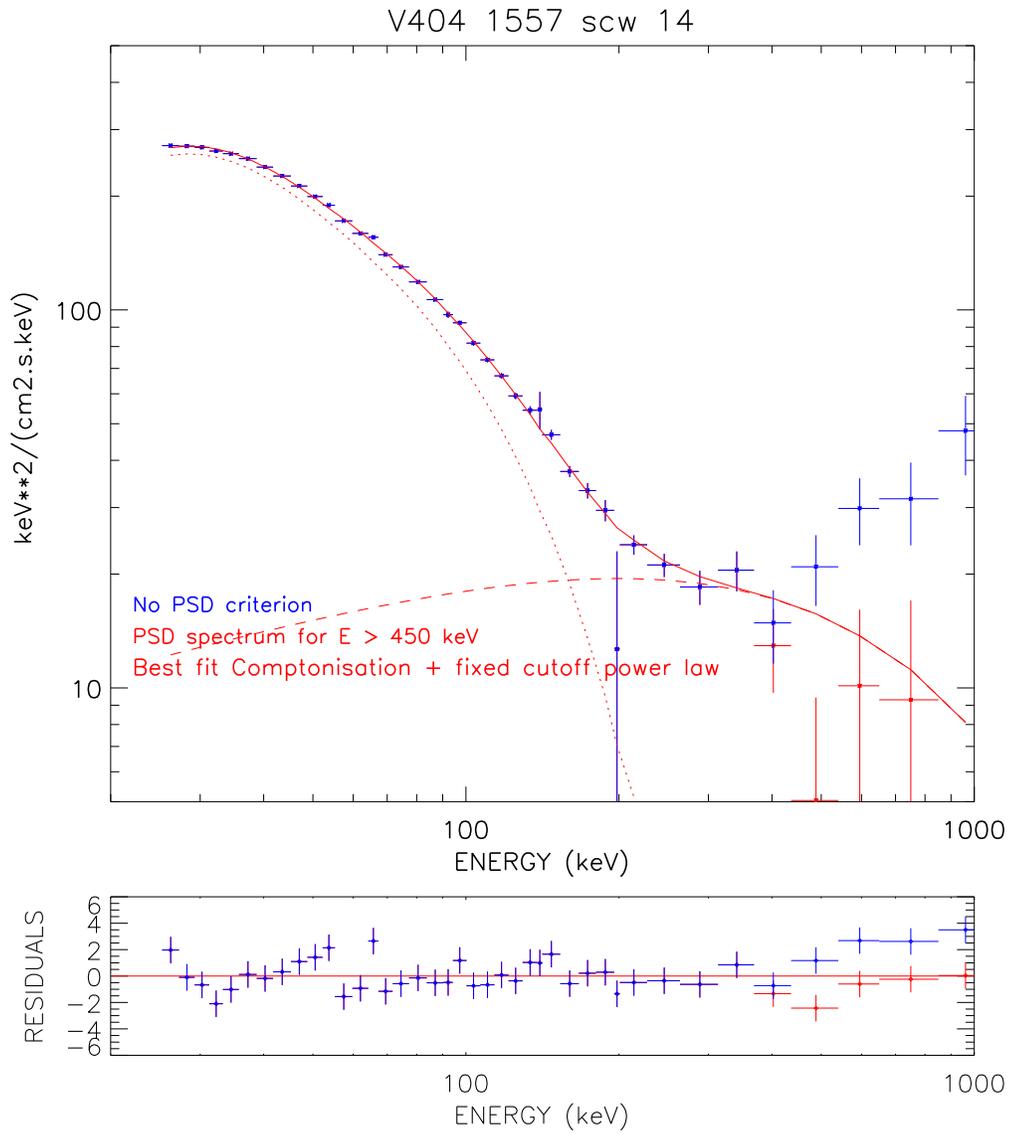}
 \caption{The same as fig. 1 for a moderate intensity flare, during revolution 1557}
\label{fig:compa1557-14}
\end{figure} 

  \begin{figure}%1
 \plotone{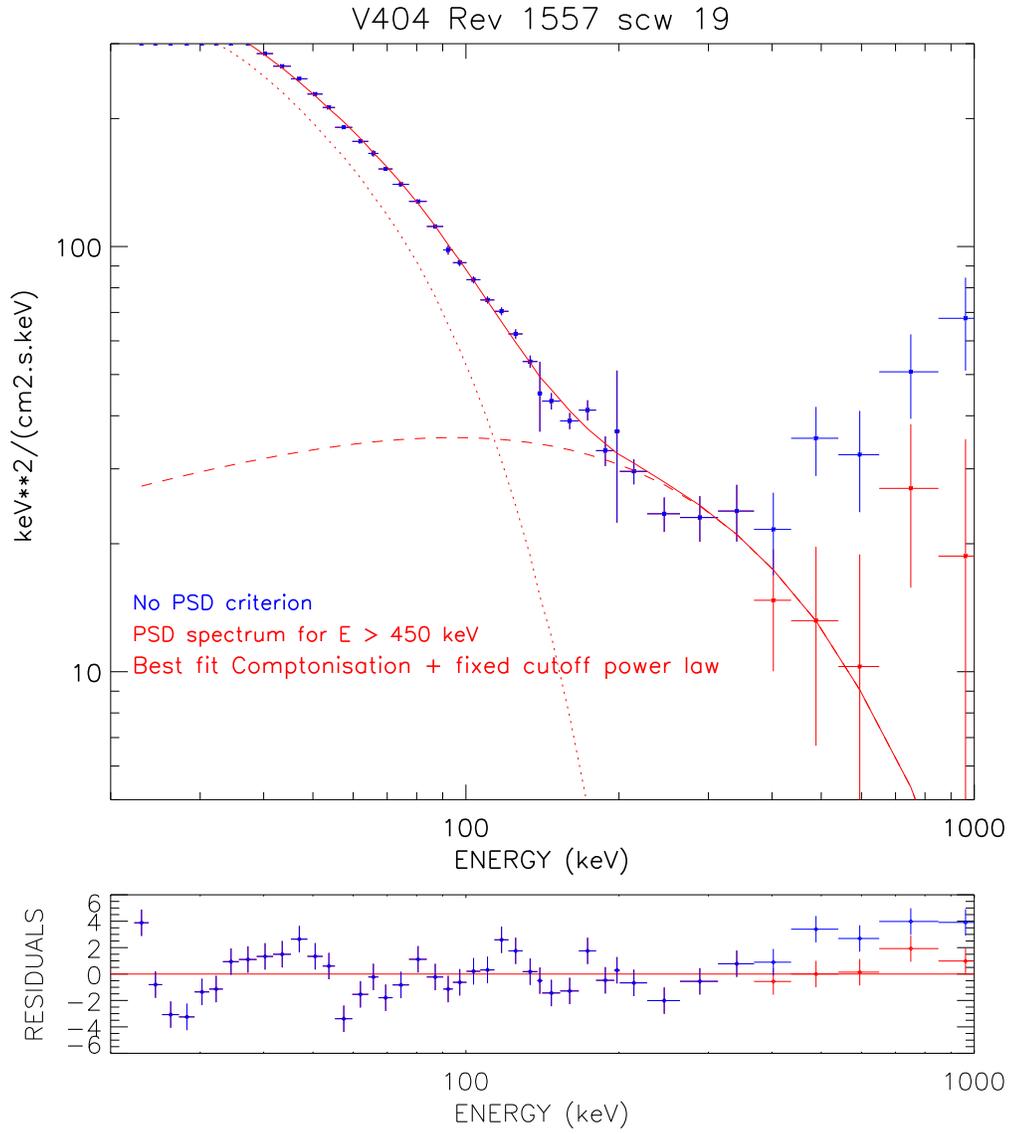}
 \caption{The same as fig. 1 for the maximum of the 2015 outburst (occuring during revolution 1557)}
\label{fig:compa1557-19}
\end{figure} 

 \begin{figure}%1
 \plotone{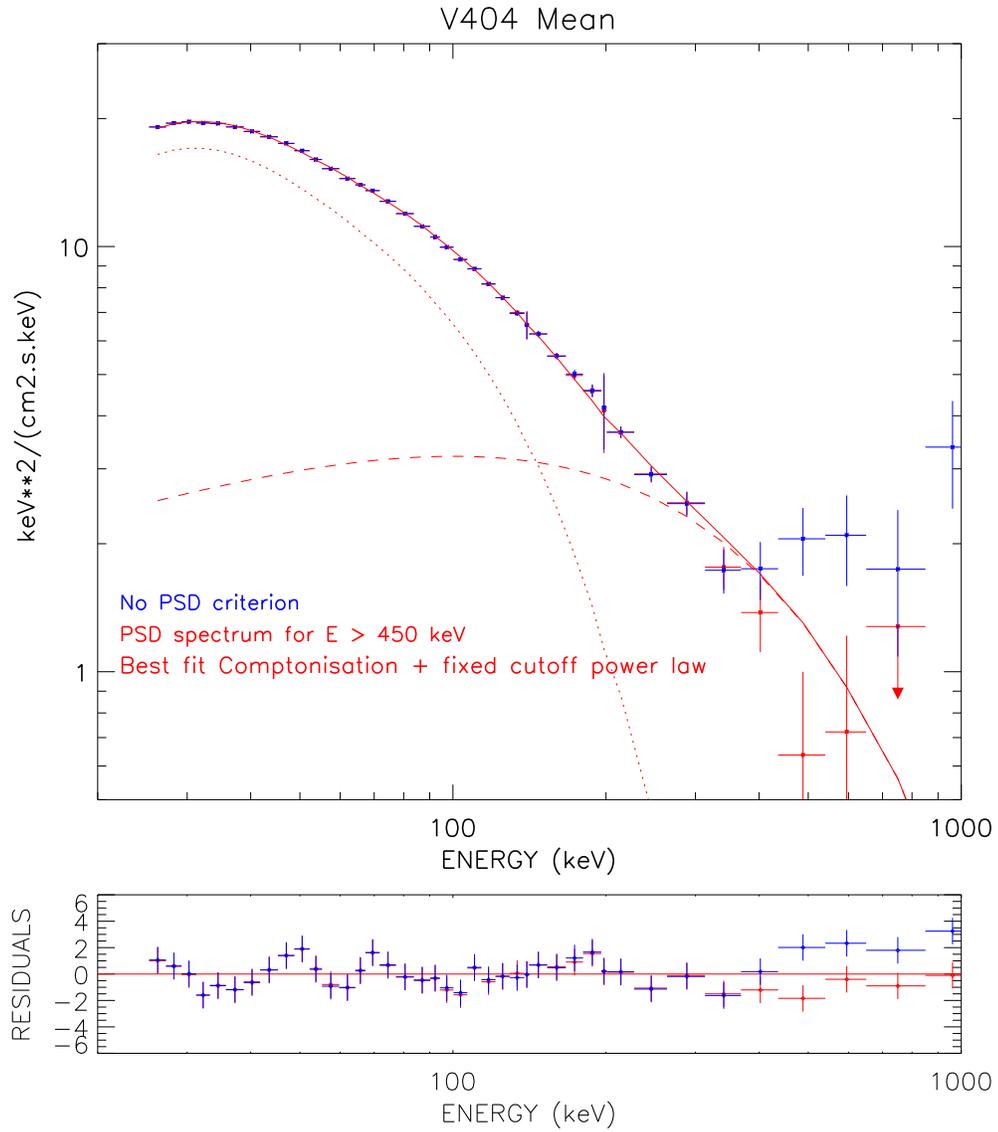}
 \caption{The same as fig. 1 for the averaged spectrum from revolution 1554 to 1559}
\label{fig:compatot}
\end{figure}

\end{document}